*In Situ* Detection of Kinetic-Size Magnetic Holes in the Martian Magnetosheath

S. Y. Huang[1*], R. T. Lin[1], Z. G. Yuan[1], K. Jiang[1], Y. Y. Wei[1], S. B. Xu[1], J. Zhang[1], Z. H. Zhang[1], Q. Y. Xiong[1], and L. Yu[1]

[1]School of Electronic Information, Wuhan University, Wuhan, 430072, China

[*]Corresponding author: shiyonghuang@whu.edu.cn

**Abstract**

Depression in magnetic field strength with a scale below one proton gyroradius is referred to as kinetic-size magnetic hole (KSMH). KSMHs are frequently observed near Earth's space environments and are thought to play an important role in electron energization and energy dissipation in space plasmas. Recently, KSMHs have been evidenced in the Venusian magnetosheath. However, observations of KSMHs in other planetary environments are still lacking. In this study, we present the *in situ* detection of KSMHs in Martian magnetosheath using Mars Atmosphere and Volatile EvolutioN (MAVEN) for the first time. The distribution of KSMHs is asymmetry in the southern-northern hemisphere and no obvious asymmetry in the dawn-dusk hemisphere. The observed KSMHs are accompanied by increases in the electron fluxes in the perpendicular direction, indicating the cues of trapped electrons and the formation of electron vortices inside KSMHs. These features are similar to the observations in the Earth's magtosheath and magnetotail plasma sheet and the Venusian magnetosheath. This implies that KSMHs are a universal magnetic structure in space.

**Introduction**

Magnetic field depressions in the total magnetic field strength are usually called magnetic holes. Magnetic holes with scales larger than a proton gyroradius, referred to as mirror modes, are frequently observed in solar wind (Zhang et al., 2008; Tsurutani et al., 2011; Volwerk et al., 2020; Yu et al., 2021), the Earth's magnetosphere (Tsurutani et al., 2011; Balikhin et al., 2009), and other planetary magnetospheres (e.g., Gattaneo et al., 1998). These large-scale magnetic holes are usually generated by ion temperature anisotropy via the mirror mode instability (Hasegawa, 1969).

Magnetic holes with scale smaller than a proton gyroradius are commonly referred to as kinetic-size magnetic holes (KSMHs). Because KSMHs can heat or accelerate electrons and dissipate energy in turbulent plasmas, KSMHs have drawn attention in recent years. KSMHs are widely observed in the Earth's magnetosphere, including magnetosheath (e.g., Huang et al., 2017a, 2017b, 2018; Yao et al., 2017; Liu et al., 2019) and magnetotail plasma sheet (e.g., Ge et al., 2011; Sundberg et al., 2015; Gershman et al., 2016; Huang et al., 2019). A variety of interesting phenomena occur inside KSMHs, including electron vortices (Huang et al., 2017a, 2017b), electron perpendicular temperature increase (Huang et al., 2017a, 2017b; Yao et al., 2017), electron-scale whistler waves, electron cyclotron waves and electrostatic solitary waves (Huang et al., 2018; Yao et al., 2019). There are several mechanisms for the generation of KSMHs, including the theory of electron solitary waves (Ji et al., 2014), electron mirror instability (Pokhotelov et al., 2013), tearing instability (Balikhin et al., 2012), and the model of electron vortex magnetic holes as coherent structures in plasma turbulence (Haynes et al., 2015; Roytershteyn et al., 2015).

Recently, Goodrich et al. (2021) presented the first evidence of KSMHs in the Venusian magnetosheath using the Parker Solar Probe. Whether KSMHs exist in other planetary environments is still unclear. In the present study, we revisit the observations from the Mars Atmosphere and Volatile EvolutioN (MAVEN) in the Martian magnetosheath and find a large number of KSMHs accompanied by electron vortices for the first time. These observations indicate that KSMHs are a universal structure in magnetized space plasmas.

**Instruments and Methods**

To investigate KSMHs, we use magnetic field data with a sampling of 32 Hz from magnetometer (MAG; *Connerney et al.*, 2015), ion data with a sampling of 4 s from the Supra-Thermal And Thermal Ion Composition (STATIC; *McFadden et al.*, 2015), and electron data with a sampling of 4 s from Solar Wind Electron Analyzer (SWEA; *Mitchell et al.*, 2016) onboard MAVEN. With a field of view of 360°×90°, STATIC is able to provide fluxes, temperature and velocity vectors of ions owing to its large mass resolution. Noted that ion density and velocity vectors are derived from STATIC measurements whereas ion temperature is derived from the MAVEN Insitu Key Parameters. SWEA measures the energy and angular distributions of the electrons from 5 eV to 5 keV in the solar wind and magnetosheath.

To identify KSMHs accurately, the following criteria are used in the present study: 1) each component of the magnetic field should remain in its direction; 2) the depression of the magnetic field **$Depth_{MH}$** should be more than 10%, where **$Depth_{MH}$**=1-$B_{min}/B_{background}$, $B_{min}$ is the minimum value of the magnetic field in the magnetic hole and $B_{background}$ is the average of the magnetic field near the magnetic hole; 3) the standard deviation of the magnetic field outside the structure should be less than that inside; 4) the length of the spacecraft crossing the magnetic hole **$L_{MH}$** should be less than one proton gyro-radius $\rho_{H+}$, where $L_{MH}=|V_{H+}|\times\Delta t$, $|V_{H+}|$ is the background proton bulk velocity and $\Delta t$ is the duration of the spacecraft passing through the magnetic hole.

**Results**

Figure 1 shows an example of KSMH on 26 Jan 2019. Figure 1a presents the energy fluxes of proton. The proton fluxes in solar wind gather around 1 keV. The energy of proton fluxes gradually expands in foreshock, and the proton fluxes visibly expand in the magnetosheath, which indicates that proton are heated crossing the bow shock. This event occurred in the Martian magnetosheath, where the proton bulk velocity is less than that in the upstream quiet solar wind (not foreshock; Figure 1c), the magnetic field is more turbulent (Figure 1e), and the density of heavy ions, such as $O^+$, remains much less than that of protons (Figure 1d). The gray shadow in the left panels of Figure 1 marks one KSMH, which is magnified in the right panels.

The MAVEN spacecraft observed a depression in magnetic field strength without reversals in the three components of the magnetic field (Figure 1h), implying the existence of a magnetic hole. MAVEN entered the magnetic hole at 10:25:32.944 UT and exited at 10:25:33.319 UT, yielding the duration $\Delta t = 0.375$ s. Based on the assumption of moving together between the magnetic hole and the plasma flow, one can estimate the size of the magnetic hole $L_{MH}$ as 80.93 km. With the temperature of background proton $T_{H+}$=141 eV, the proton gyroradius $\rho_{H+}$ is estimated to be 359 km. Thus, the size of the magnetic hole is much less than one proton gyroradius, implying that MAVEN detects one KSMH here. As shown in Figure 1f, the fluxes of electron (50-200 eV) in the perpendicular direction have significant enhancement inside this KSMH, which is similar to the observations of the KSMHs in the Earth's magnetosheath (e.g., Huang et al., 2017a, 2017b, 2018) and plasma sheet (e.g., Huang et al., 2019). It should point out that the sampling rate of SWEA is much lower than that of MAG, and the time duration of this KSMH is less than the sampling rate of SWEA. Thus, the measured fluxes of the electron belong to not only the KSMH but also one part of outside of KSMH. However, comparing the background electron fluxes, the enhancement of the electron fluxes in the perpendicular direction is obvious and can still represents the characteristic of this KSMH.

To reconstruct KSMHs, we estimate the radius $R_{MH}$ and maximum current density $J_0$ in these structures based on the simple cylindrically symmetric current vortex model proposed by Goodrich et al. (2021). The current density in the magnetic hole can be described as $\mathbf{J} = J_0 \sin(\pi r/R)$ while $r \leq R$ ($\mathbf{J}=0$, for $r > R$), where $r$ is the radial distance from the center of the magnetic hole, $R$ is the radius of the magnetic hole (i.e., the distance from the center to the outer edge of the magnetic hole), and both $J_0$ and $R$ are constant. Then, the magnetic field is obtained using Amperes Law, $\mathbf{B}_l(r)=\mathbf{B}_0-\Delta\mathbf{B}_l(r)$, where $\Delta\mathbf{B}_l(r)=\mu_0/R \int_r^R \mathbf{J}(r)r dr$. $\mathbf{B}_l(r)$ is the observed largest variation component derived from the minimum variance analysis (MVA; e.g., Sonnerup and Scheible, 1998), and $\mathbf{B}_0$ is the magnetic field $\mathbf{B}_l(r)$ at the edge of the magnetic hole (i.e., $r = R$). To obtain the best fitting results for $J_0$ and $R$, we employ the minimum mean square error method and set up a rule to restrict the minimum magnetic field in the model $\mathbf{B}_{min,model}$, which is $|\mathbf{B}_{min,model}-\mathbf{B}_{min}|/|\mathbf{B}_{min}| \leq 5\%$.

Figure 2 shows the observed magnetic field inside the KSMH shown above and the fitted result derived by the model. One can see that the fitted curve agrees well with the observed magnetic field. The estimated radius of the KSMH $R_{MH}$ is 85.80 km, which is also much smaller than $\rho_{H+}$, and the maximum current density $J_0$ is 1.10 μA/m$^2$.

According to the criteria in the previous section and the aforementioned restriction that is used in the fitted process, 102 KSMH structures are successfully selected in the Martian magnetosheath from 1 Jan 2015 to 31 Dec 2019. It should be noted that we try to present unambiguous KSMH structures that are consistent with the current vortex model. However, this process may miss some KSMH structures that are not consistent with the simple model.

Figure 3 presents the spatial distribution of all KSMH events. One can conclude several distinct features. Asymmetry exists in the southern-northern hemisphere (61.7% on the +Z axis and 38.3% on the –Z axis, Figure 3a), and no obvious asymmetry exists in the dawn-dusk hemisphere (Y axis, Figure 3b). Although several events seem to appear below the magnetic pileup boundary (MPB), the obstacle boundary that separates the magnetic pileup region from shocked solar wind, it is easy to confirm that these events are all in the magnetosheath by checking the plasma properties and the background magnetic field, which means that all identified KSMH events occur in the Martian magnetosheath. In addition, no KSMH event was observed yet in the low latitude or nose region on the dayside (Figure 3d).

Statistical studies on the parameters of KSMHs were conducted, including $L_{MH}/\rho_{H+}$, $R_{MH}/\rho_{H+}$, $Depth_{MH}$, $J_0$, $Flux_\perp/Flux_\parallel$ and $Flux_{\perp,inside}/Flux_{\perp,outside}$, where $Flux_{\perp,inside}$ (Flux

$_{\perp,\text{outside}}$) is the electron flux with a pitch angle between 60°-120° inside (outside) of the magnetic holes, and Flux$_{\parallel}$ is half of the electron flux with a pitch angle between 0°-45° and 135°-180°. Since the time resolution of SWEA is practically higher than the time duration of the KSMHs, it is almost impossible to obtain the data inside/outside of the magnetic holes precisely. Thus, the one sampling of electron fluxes covering the whole KSMH is regarded as that inside the magnetic hole, and the average of last and next sampling is that outside the magnetic hole.

Figure 4 shows the statistical results of the parameters of all KSMH events. The $L_{\text{MH}}$ normalized by $\rho_{\text{H+}}$ are mostly approximately 0.25 (Figure 4a), and the normalized estimated radius $R_{\text{MH}}$ are mostly less than 0.5 (Figure 4d). The depression of the magnetic field seems to gather at approximately 55% (Figure 4b). Most of the maximum current densities derived from the model are less than 1.2 μA/m$^2$ (Figure 4e), whereas there exist several cases with a considerably strong current (up to ~6 μA/m$^2$). The electron fluxes in the perpendicular direction inside KSMHs are higher than those outside KSMHs (the ratio could be up to 2.5, Figure 4c). Moreover, the electron fluxes in the perpendicular direction are much higher than those in the parallel direction inside KSMHs (Figure 4f). This implies that the electron fluxes in the perpendicular direction are enhanced inside KSMHs, which causes a significant increase in the electron perpendicular temperature therein.

**Discussions and Conclusions**

As mentioned in the introduction section, there are several mechanisms for the generation of KSMHs in space plasmas. However, due to the low-time resolution of both ion and electron data provided by MAVEN, one cannot directly determine which mechanisms are responsible for the KSMHs in Martian magnetosheath. According to our statistical study, the existence of KSMHs is proven in the Martian magnetosheath by MAVEN observations. All observed KSMHs are accompanied by obvious enhancements of the electron fluxes at ~90° compared to the fluxes in the parallel direction inside KSMHs and in the perpendicular direction outside KSMHs, indicating that the electrons are trapped by the KSMHs. The trapped electrons can form an electron vortex, which carries a strong current vortex to induce a magnetic field decrease inside KSMHs (e.g., Haynes et al., 2015; Huang et al., 2017a). This infers that the KSMHs in the Martian magnetosheath may be described by the electron vortex magnetic hole model. However, due to the low time resolution of the electron data, only one data point is available during the crossing of KSMHs. Thus, one cannot directly identify the electron vortex inside KSMHs through bipolar signatures in one or two components of electron velocity.

Based on the simple cylindrically symmetric current vortex model, the current and radius of the KSMHs are derived. An intense current inside KSMHs of up to 6 $\mu A/m^2$ is found. As for the case shown in Figure 1, the electron velocity is estimated as 1375 km/s using $J_0$ = 1.1 μA/m$^2$, $n_e$ ~ $n_i$ ~ 5 cm$^{-3}$, and the assumption that the currents are mainly carried by the electrons. Such a large electron velocity is much higher than the ion velocity (i.e., ~ 300 km/s) in this KSMH, indicating decoupling between the ions and the electrons. This decoupling can cause the Hall effect and lead to the Hall electric field therein.

In summary, KSMHs have been successfully identified in the Martian magnetosheath by MAVEN for the first time. These KSMHs are possibly accompanied by electron vortices, which are characterized by enhancements in electron fluxes in the perpendicular direction with respect to the ambient magnetic field. These KSMHs are likely to be explained by the electron vortex magnetic hole model in turbulent plasma. KSMHs are frequently detected in the Earth's magnetosheath and plasma sheet and have recently been detected in the Venusian magnetosheath. Our study demonstrates the existence of KSMHs in the Martian magnetosheath. Thus, KSMHs may be universal magnetic structures under different plasmas conditions in space plasma, which are indicative of a universal microphysical plasma process.

**Acknowledgement**

We thank the entire MAVEN team and instrument leads for data access and support. This work was fully supported by the National Natural Science Foundation of China

(41874191, 42074196, 41925018) and the National Youth Talent Support Program.

The data is available via https://lasp.colorado.edu/maven/sdc/public/data/.

**Figure captions**

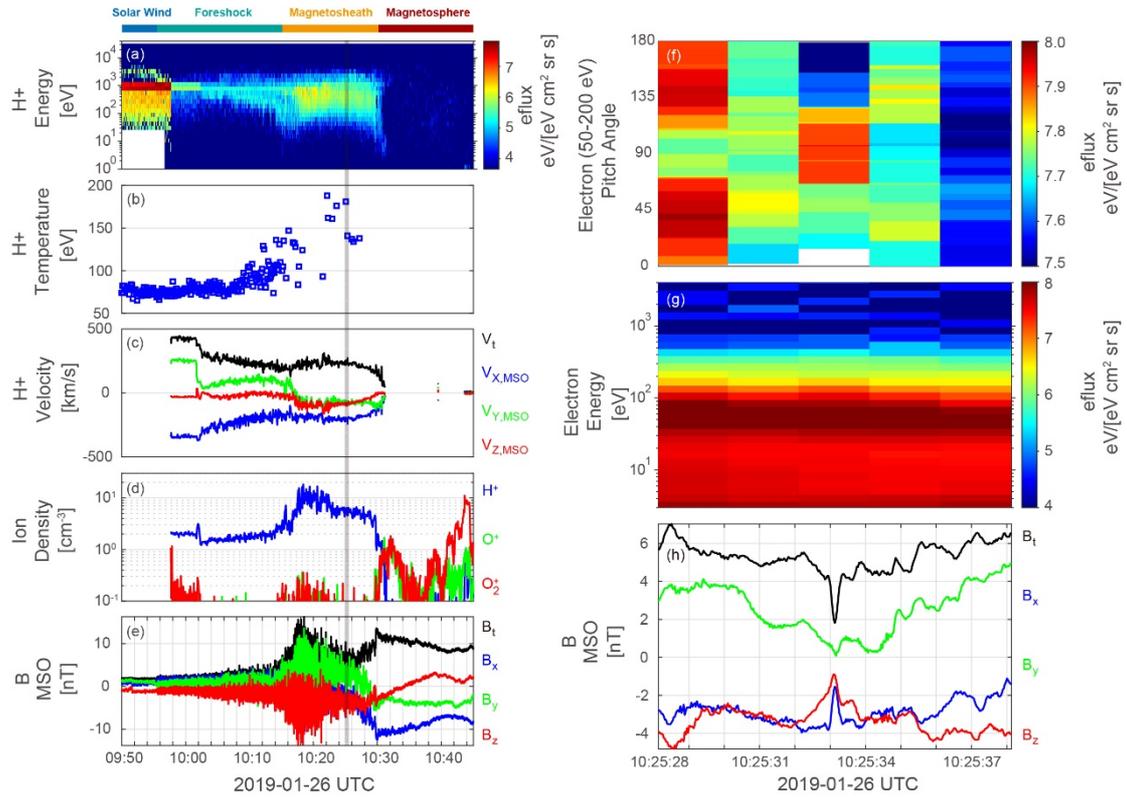

**Figure 1**. A typical example of KSMH event in the Martian magnetosheath on 26 Jan 2019. (a) proton energy spectrogram, (b) proton temperature, (c) proton velocity components in MSO coordinates, (d) ion density, (e, h) magnetic field components in MSO coordinates, (f) electron pitch angle distribution within 50-200 eV and (g) electron energy spectrogram. The shaded region indicates the time of the right panels (f-h).

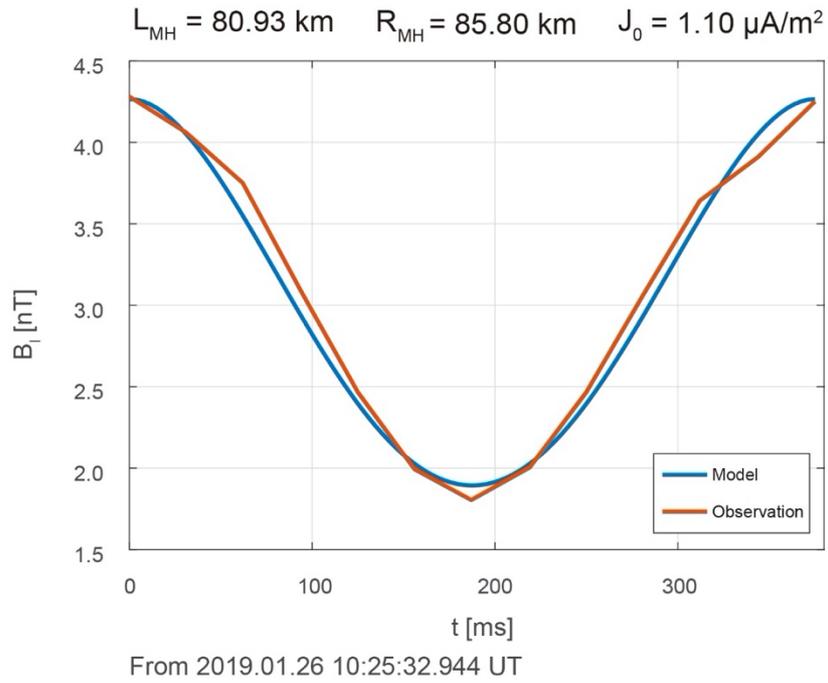

**Figure 2**. The observed magnetic field and the fitted magnetic field derived by the model. The red line represents the observed magnetic field and the blue one represents values from the model.

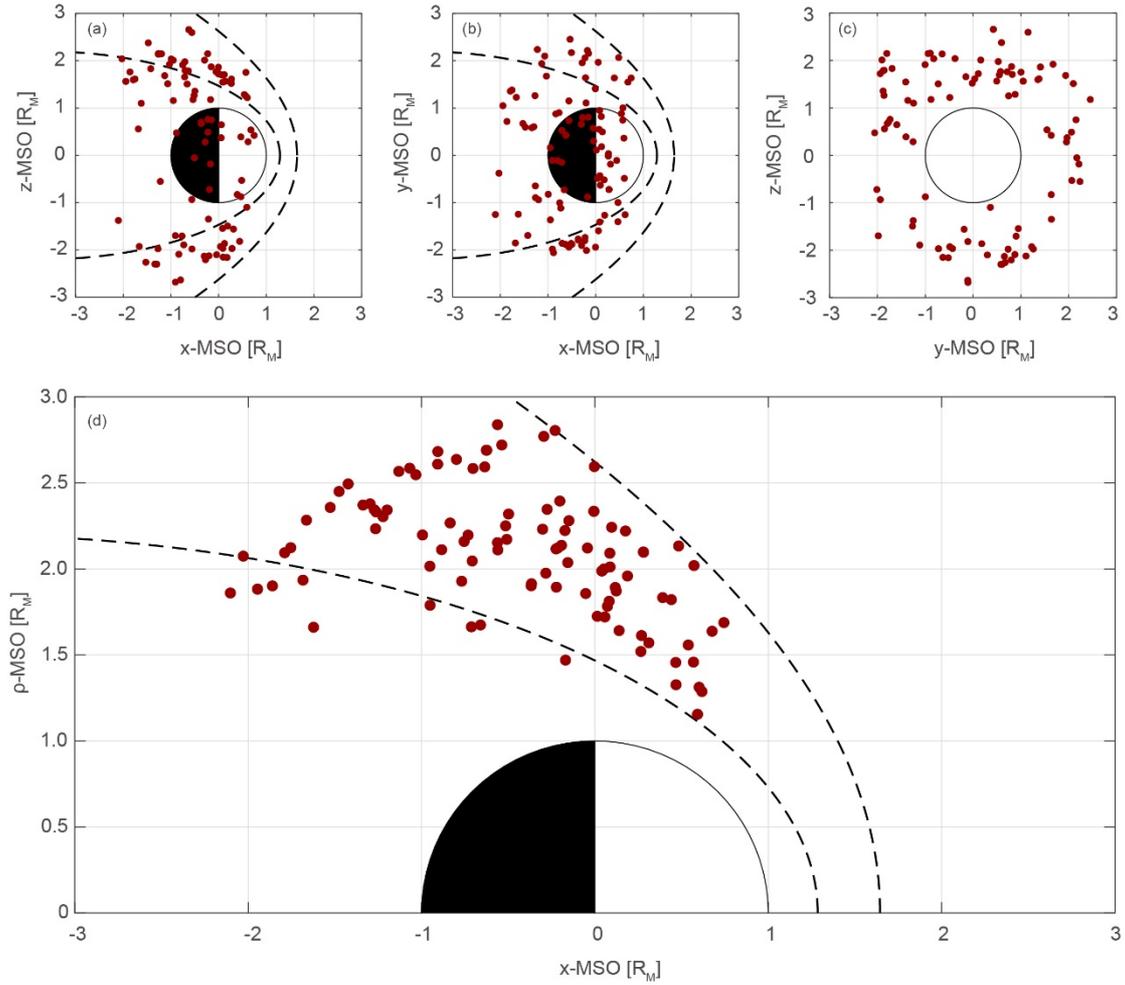

**Figure 3**. The spatial distribution of all KSMHs in Mars Solar Orbital (MSO) coordinates respectively in (a) x-z plane, (b) x-y plane, (c) y-z plane and (d) x-ρ plane, where $\rho=\sqrt{y^2 + z^2}$. The Red dots represent positions of KSMH events. Two dashed lines mark the bow shock (BS) and magnetic pileup boundary (MPB) separately (*Vignes et al.*, 2000). All axes are in units of Mars radii ($R_M$).

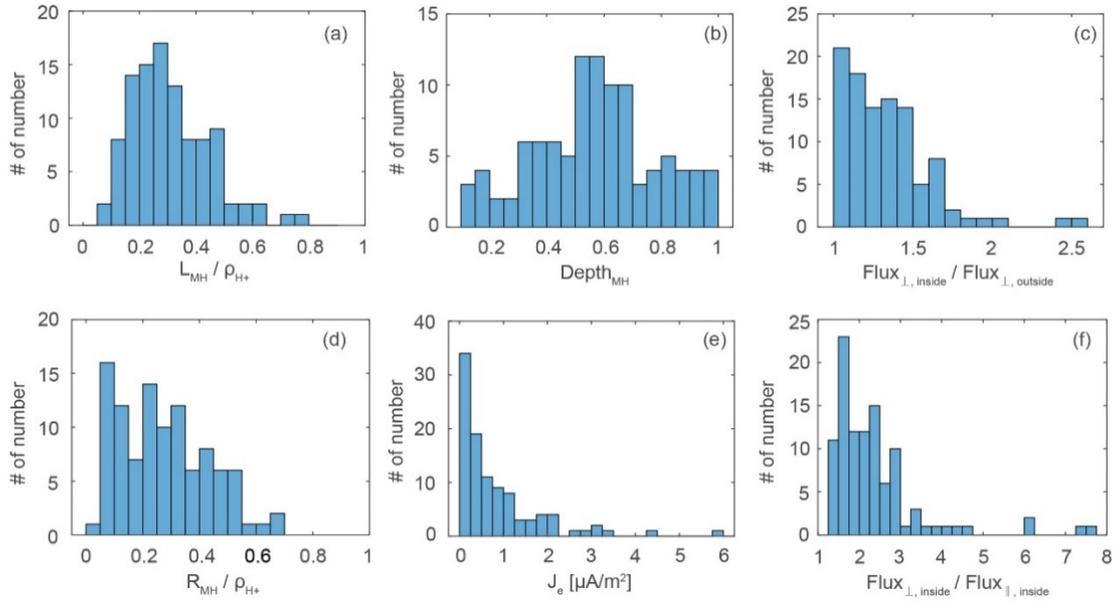

**Figure 4**. Statistical results on parameters of KSMH events. (a) $L_{MH}/\rho_{H+}$, (b) $Depth_{MH}$, (c) $\text{Flux}_{\perp,\text{inside}}/\text{Flux}_{\perp,\text{outside}}$, (d) $R_{MH}/\rho_{H+}$, (e) $J_e$ and (f) $\text{Flux}_{\perp,\text{inside}}/\text{Flux}_{\parallel,\text{inside}}$ inside KSMHs. $L_{MH}$ is the length of the spacecraft crossing the KSMHs, $\rho_{H+}$ is the proton gyroradius, $R_{MH}$ is the estimated radius, $Depth_{MH}$ is the depression of magnetic field, $J_e$ is maximum current density derived from the model, $\text{Flux}_{\perp,\text{inside}}$ ($\text{Flux}_{\perp,\text{outside}}$) is the electron fluxes with pitch angle between 60°-120° inside (outside) of the structure and $\text{Flux}_{\parallel}$ is half of the electron fluxes with pitch angle between 0°-45° and 135°-180°.